\documentstyle[12pt,preprint,tighten,floats,aps,epsf,psfig]{revtex}
\begin{document}
\title{Interacting electrons in a 2D quantum dot}
\author{N.Akman\footnote{e-mail:akman@newton.physics.metu.edu.tr}, 
M.Tomak\footnote{e-mail:tomak@rorqual.cc.metu.edu.tr}}
\date{{\small Middle East Technical University, Department of Physics, 
06531 Ankara, Turkey}}
\maketitle
\begin{abstract}
The exact numerical diagonalization of the Hamiltonian of a 2D circular 
quantum dot is performed for 2, 3, and 4 electrons.The results are 
compared with those of the perturbation theory.Our numerical results
agree reasonably well for small values of the dimensionles coupling 
constant $\lambda=\frac{a}{a_{B}}$ where $a$ is the dot radius and $a_{B}$
is the effective Bohr radius.Exact diagonalization results are compared
with the classical predictions, and they are found to be almost
coincident for large $\lambda$ values.  
\end{abstract}
\newpage
\section{Introduction}
There is considerable interest in the basic science and technological
applications of quantum dots [1-5].The electron motion in quantum dots 
are confined to a region with dimensions comparable to the de Broglie
wavelength of the particle.The result is the quantization of energy. 
However, since the quantization in the vertical direction is 
much stronger than in the planar directions, a quantum dot can well be 
treated as a 2D disc of finite radius.

Although quantum dots and natural atoms contain comparable
number of electrons, their electronic properties can have
gross differences due to the nature of the confinement
potential [6].In quantum dots confinement potential
is experimentally controllable [4], and usually parabolic
and much shallower than the $\frac{1}{r}$ potential of 
the natural atoms.One recalls, however, that parabolic structures
are not the only ones because hard wall type confinement
can also be obtained by etching techniques [7-10].
Electron electron correlations in quantum dots have been
worked out for parabolic and hard wall confinements
in the classical limit and finite temperature in [11],
and for hard wall confinement at zero temperature in [12]. 

In this work we perform an exact numerical analysis
of a few electrons confined in a circular 2D quantum dot.
We believe that such analyses of 2D structures can shed light
on the nature of the electron correlations in quantum dots.

The organization of the paper is as follows.In Sec. 2 we formulate
the second quantized Hamiltonian and specify its symmetries.
In Sec. 3 we perform a numerical diagonalization of the Hamiltonian
obtained its energy levels as a function of the dot dimension. 
We also present a comparative analysis of the quantum mechanical and 
classical energy levels.
\section{Model and the method of calculation}
We consider a many electron system in a two-dimensional circular 
quantum dot with radius $a$.The system is described by the Hamiltonian
\begin{eqnarray}
H=\sum_{i=1}^{N}\{-\frac{\hbar^{2}}{2m^{\star}}{\bf \vec{\nabla}}_{i}^{2}+
   U({\bf \vec{x}} _{i})\}+
  \frac{e^{2}}{2\epsilon}\sum_{i \neq j}\frac{1}{|{\bf \vec{x}} _{i}-{\bf \vec{x}} _{j}|}\,\,, 
\end{eqnarray}
where $m^{\star}$ is the electron effective mass and  
$\epsilon$ is the dielectric constant of the medium.The 
second term in the paranthesis, $U({\bf \vec{x}} )$, is the hard-wall confinement
potential which vanishes for $|{\bf \vec{x}} |<a$, and infinite for 
$|{\bf \vec{x}} |\geq a$.For convenience, we introduce the dimensionless 
Hamiltonian ${\cal{H}}$ via the definition
\begin{eqnarray}
H=\frac{\hbar^{2}}{m^{\star}a_{B}^{2}\lambda^{2}}{\cal{H}},
\end{eqnarray}
where $\lambda=a/a_{B}$ is the dimensionless coupling constant.The 
dimensionless Hamiltonian ${\cal{H}}$ is given by
\begin{eqnarray}
{\cal{H}}=\sum_{i=1}^{N}(-\frac{1}{2}){\bf \vec{\nabla}}_{i}^{2}+
           \frac{\lambda}{2}\sum_{i \neq 
           j}\frac{1}{|{\bf \vec{x}} _{i}-{\bf \vec{x}} _{j}|}.
\end{eqnarray}
One notes that energy is measured in units of 
$\frac{\hbar^{2}}{m^{\star}a_{B}^{2}\lambda^{2}}$ which depends on $\lambda$,
and length is measured in units of $a$, the dot radius.

In the second quantized language, the dimensionless Hamiltonian takes 
the form
\begin{eqnarray}
{\cal{H}}=\sum_{K}{\cal{E}}_{K}a_{K}^{\dagger}a_{K}+\frac{\lambda}{2}
          \sum_{K,L,M,N}
          V_{K,L,M,N}\;a_{K}^{\dagger}a_{M}^{\dagger}a_{L}a_{N}\,\,,
\end{eqnarray}
where ${\cal{E}}_{K}$ is the single particle energy level, and
\begin{eqnarray}
V_{K,L,M,N}= \int \int d^{2}{\bf \vec{x}}  d^{2}{\bf \vec{x}} ^{\prime} 
             \varphi_{K}^{\star}({\bf \vec{x}} )
             \varphi_{L}({\bf \vec{x}} )V({\bf \vec{x}} -{\bf \vec{x}} ^{\prime})
             \varphi_{M}^{\star}({\bf \vec{x}} ^{\prime})
             \varphi_{N}({\bf \vec{x}} ^{\prime})
\end{eqnarray}
is the matrix element of the Coulombic interaction term in (3).Here
$\varphi_{A}({\bf \vec{x}} )$ ($A=K, L, M, N$) is the single particle eigenstate
of the free Hamiltonian, and $A$ is a collective index designating the radial
($n_{A}$), angular momentum ($m_{A}$) and spin ($\sigma_{A}$) quantum 
numbers.Solution of the eigenvalue problem for the free Hamiltonian yields 
\begin{eqnarray}
\varphi_{A}({\bf \vec{x}} )=\phi_{n_{A},m_{A}}({\bf \vec{x}} )\chi_{\sigma_{A}}\,\,,
\end{eqnarray}
where $\chi_{\sigma}$ is the spin wavefunction, and 
$\phi_{n,m}({\bf \vec{x}} )$ is the normalized orbital eigenfunction given by  
\begin{eqnarray}
\phi_{n,m}({\bf \vec{x}} )=\frac{1}{\sqrt{\pi}}\frac{1}{|J_{|m|+1}(k_{n,|m|})|}
                     e^{im\theta}J_{|m|}(k_{n,|m|}|{\bf \vec{x}} |)\,\,,
\end{eqnarray}
with the eigenvalues ${\cal{E}}_{K}=\frac{1}{2}k_{n_{K},|m_{K}|}^{2}$ 
which is independent of both spin and the sign of $m$. 
Here $k_{n,|m|}$ are the 
zeroes of the Bessel function ($J_{|m|}(k_{n,|m|})=0$).

In determining the spectrum of the total Hamiltonian in (3), it would be 
convenient to find a representation for the Coulombic potential
$V({\bf \vec{x}} -{\bf \vec{x}} ^{\prime})=\frac{1}{|{\bf \vec{x}} -{\bf 
\vec{x}} ^{\prime}|}$ in terms of Bessel functions, as the
single particle eigenstates in (7) do depend only on the Bessel functions.
Accordingly, we use the following decomposition for
$1/|{\bf \vec{x}} -{\bf \vec{x}} ^{\prime}|$
\begin{eqnarray}
\frac{1}{|{\bf \vec{x}} -{\bf \vec{x}} ^{\prime}|}=\sum_{m=-\infty}^{\infty}
             \int_{0}^{\infty}dk\,e^{im(\theta-\theta^{\prime})} 
             J_{m}(k\rho)J_{m}(k\rho^{\prime})e^{-k(z_{>}-z_{<})},
\end{eqnarray} 
where we let $z_{>}-z_{<}\rightarrow 0$ as the vertical dimension
of the dot is vanishingly small.

By inserting the equations (6), (7) and (8) into equation (5), we 
obtain the final form of the spin independent potential matrix element  
$V_{K,L,M,N}$ as follows
\begin{eqnarray}
&&V_{m_{K},m_{L},m_{M},m_{N}}^{n_{K},n_{L},n_{M},n_{N}}=
4\frac{1}{|J_{|m_{K}|+1}(k_{n_{K},|m_{K}|})|}
\frac{1}{|J_{|m_{L}|+1}(k_{n_{L},|m_{L}|})|}
\frac{1}{|J_{|m_{M}|+1}(k_{n_{M},|m_{M}|})|}\nonumber\\
&&\frac{1}{|J_{|m_{N}|+1}(k_{n_{N},|m_{N}|})|}
\int_{0}^{\infty}dk\,\int_{0}^{1} d\rho\, \rho  
J_{|m_{K}|}(k_{n_{K},|m_{K}|}\rho)
J_{|m_{L}|}(k_{n_{L},|m_{L}|}\rho)\nonumber\\ 
&&J_{m_{K}-m_{L}}(k\rho) \int_{0}^{1} d\rho^{\prime}\, \rho^{\prime} 
J_{|m_{M}|}(k_{n_{M},|m_{M}|}\rho^{\prime})
J_{|m_{N}|}(k_{n_{N},|m_{N}|}\rho^{\prime})
J_{m_{N}-m_{M}}(k\rho^{\prime})\,\,.
\end{eqnarray}

Integration over the azimuthal angle $\theta$ in (5) drops the sum over 
$m$ in (8),
and the difference between the orbital angular momentum quantum numbers 
becomes the
order of the Bessel function.In (9) $J_{m_{K}-m_{L}}$ and $J_{m_{N}-m_{M}}$ 
arise
from such angular integrations, and unlike the other Bessel functions in 
(9), they have negative or positive order index.
As $J_{-|m_{K}-m_{L}|}=(-1)^{m_{K}-m_{L}}J_{|m_{K}-m_{L}|}$ and
$J_{-|m_{N}-m_{M}|}=(-1)^{m_{N}-m_{M}}J_{|m_{N}-m_{M}|}$,
these two Bessel functions create sign differences among different
potential matrix elements.

From the symmetries of the problem it is seen that the $z$ component of the
total spin $S_{z}=\sum_{i=1}^{N}s_{i}$ and the $z$ component of the 
orbital angular momentum $M=\sum_{i=1}^{N}m_{i}$ are conserved quantum 
numbers.The diagonalization of the Hamiltonian is done in the single 
particle basis by taking the constraints coming from the conservation of 
z component of total spin and orbital angular momentum.Hence, dimension 
of the total Hamiltonian matrix depends on $S_{z}$ and $M$.
In the next section we shall present the result of the exact diagonalization
of the many body Hamiltonian matrix for 2, 3 and 4 electrons.
\section{Numerical results and discussion}
In this section we present the $\lambda$ dependence of the ground state 
energies for 2, 3 and 4 electrons.For each electron number we consider 
certain values of $S_{z}$ and $M$.In the figures below we plot the ground 
state energy of the dimensionless Hamiltonian ${\cal{H}}$ 
in (3) as a function of $\lambda$.

The dimensionless coupling constant $\lambda$ measures the dot dimension 
in units of Bohr radius.In plotting the figures we shall vary $\lambda$ from
0 to 3.For $0\leq \lambda <1$ one can analyze this problem using 
perturbation theory as was already done in [13].However when $\lambda$ 
exceeds 
unity perturbation theory methods do not work, and one has to resort to some
other technique, such as numerical diagonalization.Hence, ground state energies
will be obtained by the exact diagonalization of the dimensionless many 
body Hamiltonian.

In Fig.1 depicted is the $\lambda$ dependence of the ground state energy
for 2 electrons.Here solid curve is for $2S1$ ($M=0$, $S_{z}=0$),
dashed curve for  
$2S2$ ($M=1$, $S_{z}=0$), and short-dashed curve for $2T1$
($M=1$, $S_{z}=1$).

Fig.2 displays the $\lambda$ dependence of the ground state energy for 3
electrons.In this figure curves correspond to the ground state energies of
$3D1$ ($|M|=1$, $S_{z}=1/2$) (solid curve), $3D3$ ($|M|=0$, $S_{z}=1/2$)
(short-dashed curve),
$3D2$ ($|M|=2$, $S_{z}=1/2$) (dashed curve), and $3Q1$ ($|M|=0$, $S_{z}=3/2$)
(dotted curve).

In Fig.3 we present the case of 4 electrons.The solid curve is for
$4S1$ ($M=2$, $S_{z}=0$), dashed curve for $4S2$ ($M=0$, $S_{z}=0$),
short-dashed curve for $4S3$ ($M=1$, $S_{z}=0$), dotted curve for 
$4T1$ ($M=0$, $S_{z}=1$), and dot-dashed curve for
$4T2$ ($M=1$, $S_{z}=1$).

As is seen in these three figures, our exact diagonalization results
are comparable with those ones which are obtained from perturbation
theory [13].However, there are some differences between our numerical 
results and perturbative results.For instance, $2S2$ state in Fig.1 and 
$3D3$ state in Fig.2 are
energetically more favoured than $2T1$ and $3Q1$ states, respectively.
Also in Fig.3, $4S2$ and $4S3$ states are more stable than $4T1$ and
$4T2$ respectively.

It might be interesting to compare the exact diagonalization results
with those of the classical considerations.Modelling the quantum dot by an
isolated
conducting disc with the capacitance $C=(2\epsilon a)/\pi$ one gets
the classical interaction energy $E_{int.}^{class.}=\frac{1}{\lambda}$,
in units of
$\frac{\hbar^{2}}{m^{\star}a_{B}^{2}}\frac{\pi N^{2}}{4}$,
where $N$ is the number of electrons.In Fig.4 we present the $\lambda$ 
dependence of the quantum
mechanical interaction
energies for $2S1$ (solid curve), $3D1$ (dashed curve) and $4S2$ (short
dashed curve) in units of
$\frac{\hbar^{2}}{m^{\star}a_{B}^{2}}\frac{\pi N^{2}}{4}$.
The dotted curve in this figure corresponds to the classical
interaction energy.
In fact, we see that exact diagonalization results always remain
below the classical result [13], and approach to it for large
$\lambda$ values.One further observes that for small $\lambda$ values,
quantum mechanical and classical results behave differently, as
opposed to the case of large $\lambda$.
This can be attributed to the effect of electron-electron correlations,
which implies that the ground state energy cannot be written as the sum of
the contributions of kinetic energy and interaction potential [12].

Figure 5 shows the dependence of the dimensionless ground state energy
per particle, ${\cal{E}}_{0}/N$, on the particle number $N$ for different
$\lambda$ values.For small $\lambda$ values, ${\cal{E}}_{0}/N$ deviates 
from a linear
behaviour due to the dominance of the free Hamiltonian.On the other hand,
when the interaction potential becomes dominant with increasing $\lambda$
value, ${\cal{E}}_{0}/N$ changes almost linearly with $N$, as is observed
in the figure. 

In conclusion, we have worked out the exact diagonalization of the total
Hamiltonian for 2, 3 and 4 electrons in a 2D circular quantum dot
as a function of the dot dimension.Our results approximately agree
with those of the perturbation theory for small radii, but disaggrement
sets in for larger radii.We have also compared exact diagonalization
results with the classical predictions, and found that they almost coincide
for large $\lambda$ values.

\newpage
\section{Figure Captions}
Figure 1: The ground state energy levels for 2 electrons\\
Figure 2: The ground state energy levels for 3 electrons\\
Figure 3: The ground state energy levels for 4 electrons\\
Figure 4: Comparison of the classical interaction energy with the lowest 
three quantum mechanical interaction energies\\ 
Figure 5: Dimensionless ground state energies per particle
${\cal{E}}_{0}/N$ versus the particle number $N$ for $\lambda=5(\Diamond)$,
$\lambda=10(+)$, $\lambda=50(\Box)$, $\lambda=100(\times)$\\
\newpage
\begin{figure}
\vspace{15 cm}
\end{figure}
\begin{figure}
\vspace{5.0cm}
    \includegraphics{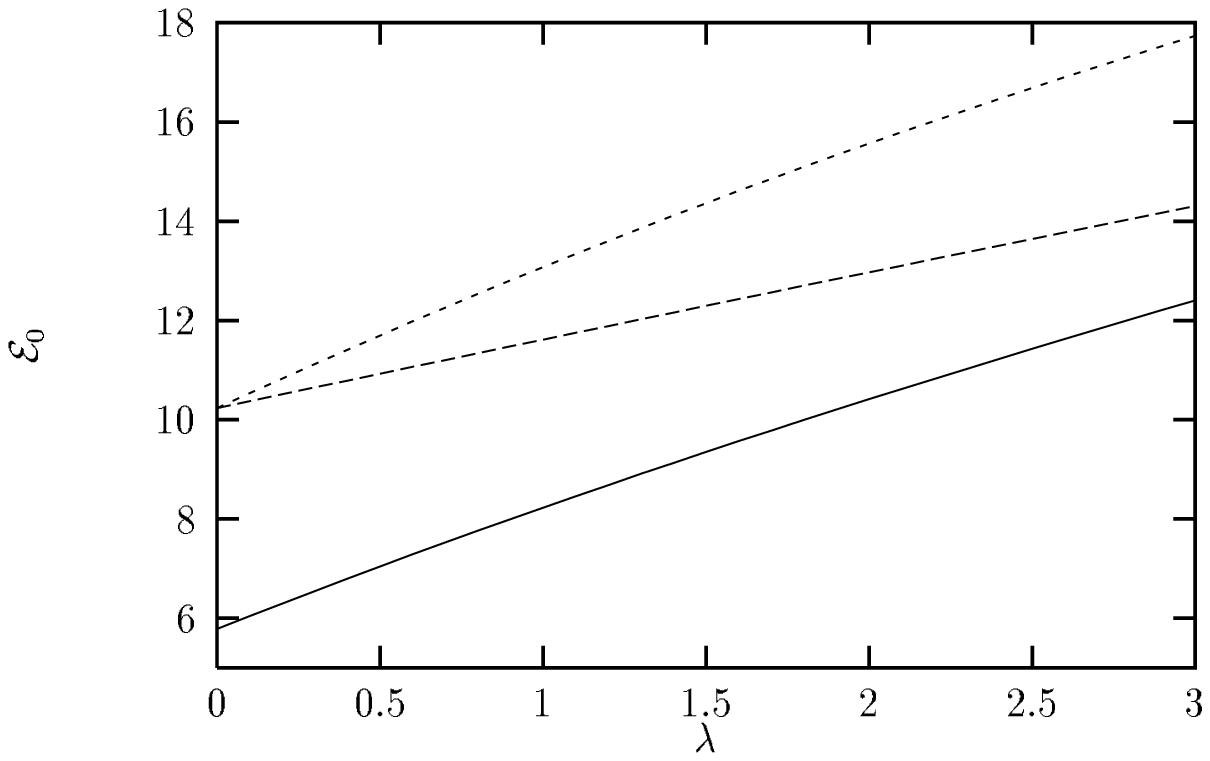}  
\mbox{\bf{Fig.1}}
\end{figure}
\begin{figure}
\vspace{7.0cm}
\end{figure}  
\begin{figure}
\vspace{12.0cm}
    \includegraphics{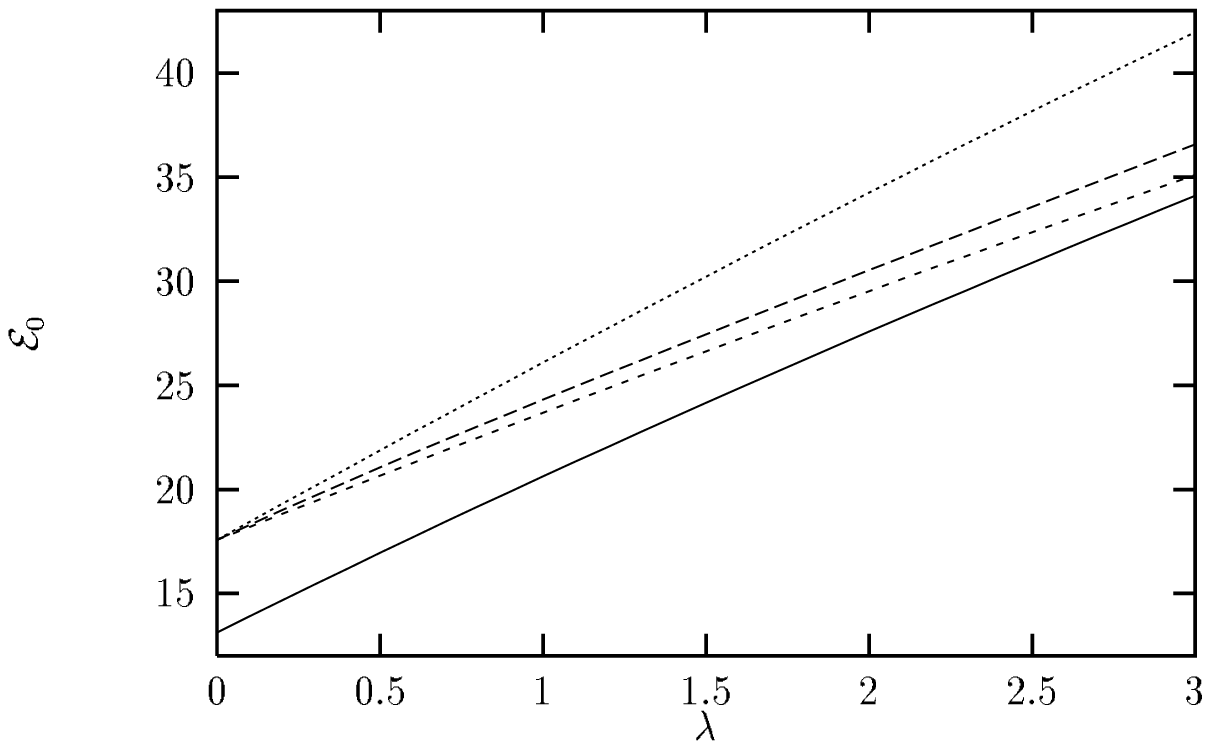}   
\mbox{\bf{Fig.2}}
\end{figure}   
\begin{figure}
\vspace{8.0cm} 
\end{figure}
\begin{figure}
\vspace{12.0cm}
    \includegraphics{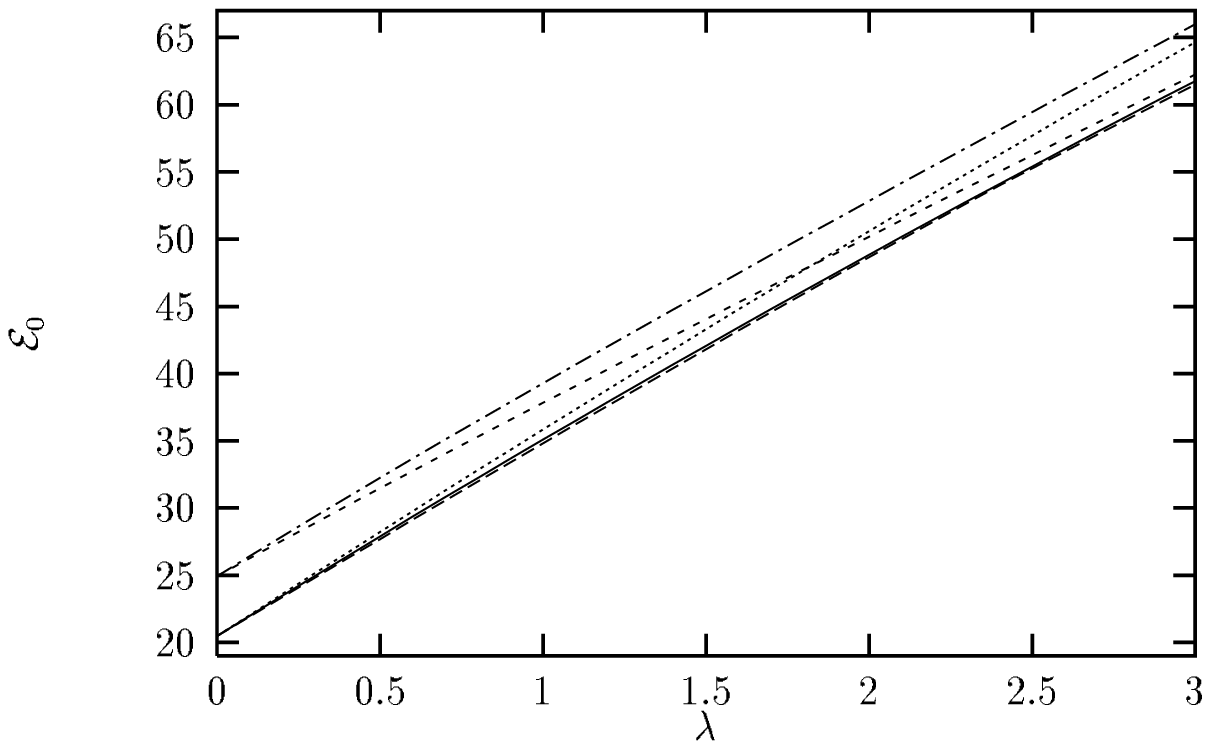}  
\mbox{\bf{Fig.3}}
\end{figure}  
\begin{figure}
\vspace{7.0cm}
\end{figure}  
\begin{figure}
\vspace{12.0cm}
    \includegraphics{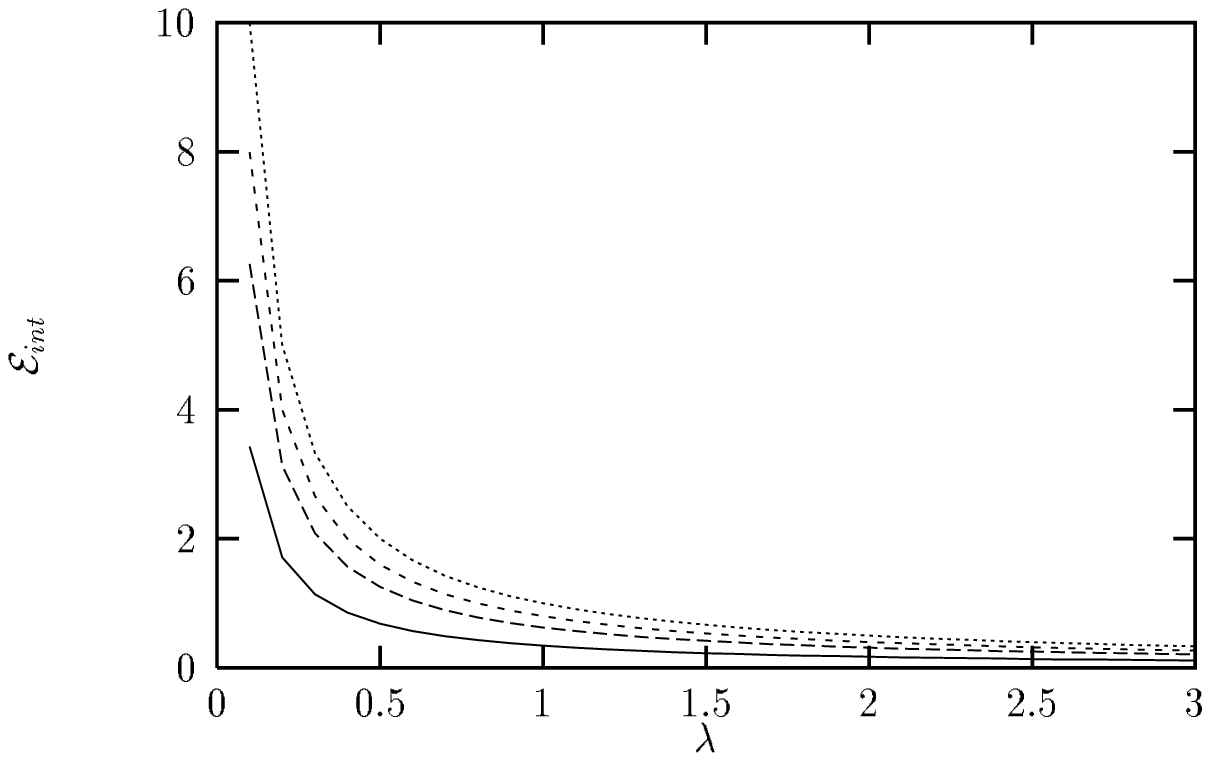}
\mbox{\bf{Fig.4}}
\end{figure}
\begin{figure}
\vspace{7.0cm}
\end{figure}  
\begin{figure}
\vspace{12.0cm}
    \includegraphics{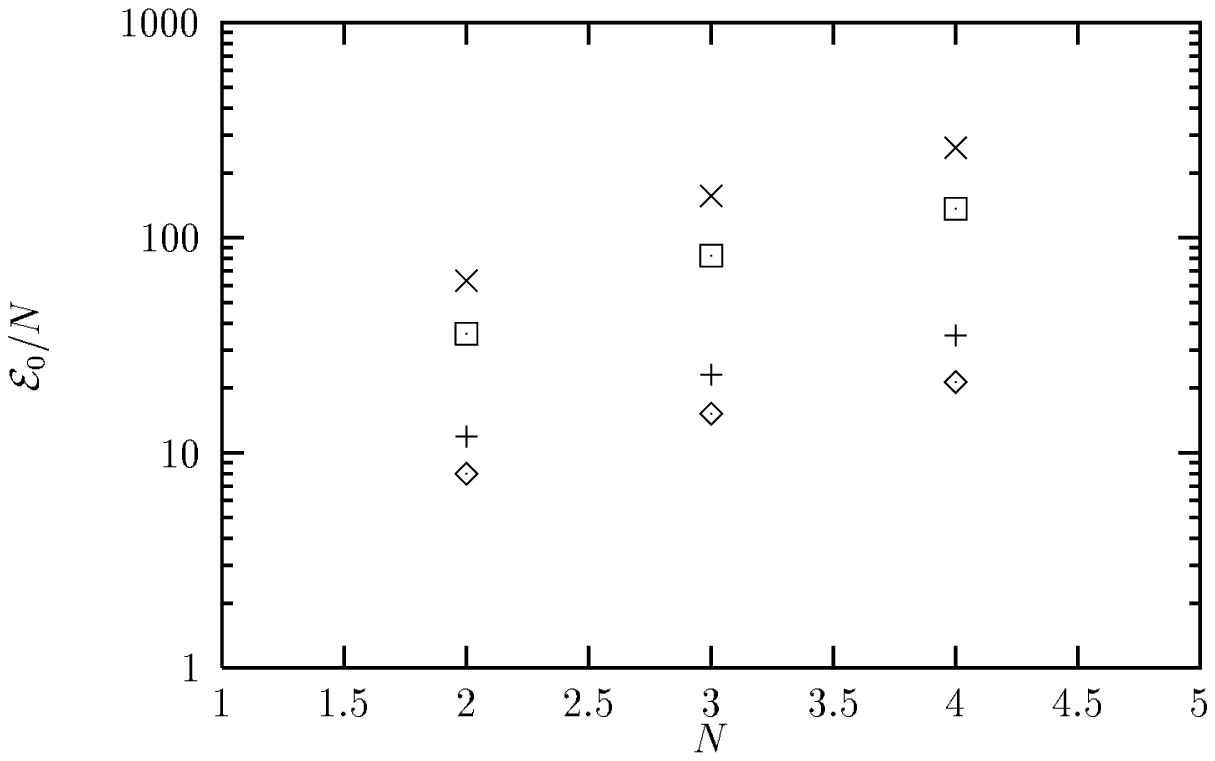}
\mbox{\bf{Fig.5}}  
\end{figure}

\begin{thebibliography}{99}
\bibitem{1} {Fulton, T.A. and Dolan, G.J., Phys. Rev. Letters 
{\bf 59}, 109 (1987).}
\bibitem{2} {Meirav, U. et al, Z. Phys. {\bf B85}, 357 (1991).}
\bibitem{3} {Kastner, M., Rev. Mod. Phys. {\bf 64}, 849 (1992).}
\bibitem{4} {Kastner, M., Physics Today, {\bf January}, 24 (1993).}
\bibitem{5} {" Single Charge Tunneling " (Editted by H.Grabert and 
M.H. Devoret) (NATO ASI Series, Plenum Press, 1992), vol. 294.} 
\bibitem{6} {Jefferson, J.H. and Hausler, W., Cond-Mat 9705012.}
\bibitem{7} {Reed, M.A. et al, Phys. Rev. Lett. {\bf 60}, 535 (1988).}
\bibitem{8} {Leonard, D., Krishnamurthy, M., Reaves, C.M., 
Denbaars, S.P. and Petroff, P.M., Appl. Phys. Lett. {\bf 63}, 
3203 (1993).}
\bibitem{9} {Notzel, R., Fukui, T. and Hasegawa, H., Appl. Phys. 
Lett. {\bf 65}, 2854 (1994).}
\bibitem{10} {Oshinowo, J., Nishioka, M., Ishida, S. and Arakawa, Y.,
Appl. Phys. Lett. {\bf 65}, 1421 (1994).}
\bibitem{11} {Bedanov, V.M., Peetes, F.M., Phys. Rev. {\bf B49}, 
2667 (1994).}
\bibitem{12} {Hausler, W. and Kramer, B., Phys. Rev. {\bf B47}, 
16353 (1993).}
\bibitem{13} {Fjarestad, J.O., Matulis, A. and Chao, K.A., Physica 
Scripta {\bf T69}, 138 (1997).}
\end{thebibliography}
\end{document}